\documentclass{article}  

\usepackage[scaled]{helvet} % ss
\usepackage{courier} % tt
\usepackage{eulervm} % a better implementation of the euler package (not in gwTeX)
\normalfont
\usepackage[T1]{fontenc}

\usepackage{pict2e}
\usepackage{mathrsfs}
\usepackage{hyperref}
\usepackage{amssymb,amsmath,amsthm}
\usepackage{bbold}
\usepackage{color}

\newcommand{\beq}{\begin{equation}}
\newcommand{\eeq}{\end{equation}}

\begin{document}

\title{Canonical Interacting Quantum Fields on Two-Dimensional De Sitter Space}

%\pacs{81T08, 82B21, 82B31, 46L55} 

%\bigskip
%\keywords{Constructive quantum field theory, Noether theorem, stress-energy tensor, 
%geodesic KMS condition, de Sitter space, quantum field theory on curved space-time.}

\author{Christian D.\ J\"akel\footnote{jaekel@ime.usp.br, Universidade de S\~ao Paulo (USP), Brasil} \ 
and Jens Mund\footnote{mund@fisica.ufjf.br, Departamento de Fisica, Universidade de Juiz de Fora, Brasil}}        

\maketitle

\begin{abstract}
We present the ${\mathscr P}(\varphi)_2$ model on de Sitter space in the canonical formulation. 
We discuss the role of the Noether theorem and we provide explicit expressions for 
the energy-stress tensor of the interacting model. 
\end{abstract}

\section{Introduction and Summary}
\label{sec_1}

The symmetry group of de Sitter space is the \emph{Lorentz group}, so one might expect that 
one can easily define a \emph{relativistic} quantum field theory on de Sitter background, 
which shares the basic characteristics with its Minkowski space counterpart. 
But there are substantial conceptual and technical obstacles, if one wants to 
formulate an \emph{interacting} quantum field theory on a (general) curved space-time. If the space-time 
is maximally symmetric (as is the case for de Sitter space), some of them can be overcome
but others are so severe that they essentially hindered much progress. Lorentzian perturbation theory is particularly effected, as 
its conceptual basis is invalidated:  
de Sitter space does not allow a non-vanishing globally time-like Killing vector field. Hence,  
there is \emph{no natural notion of energy}, \emph{no mass gap} in the energy-momentum spectrum (in fact, the latter is not
even defined) 
and \emph{no isolated mass-shells} which could be 
used in a perturbative approach. As a consequence, one is confronted with severe infrared 
singularities, despite the fact that these should be absent~\footnote{Due to an unfortunate choice of 
coordinates (which was adopted due to the formulation of Osterwalder-Schrader positivity used in \cite{FHN})  
the absence of infrared singularities was not visible in the 
non-perturbative treatment of the $\mathscr{P}(\varphi)_2$ model
by Figari, H\o egh-Krohn and Nappi \cite{FHN} published in 1975.} in a non-perturbative approach due to the finite spatial 
volume of de Sitter space. 

The situation is slightly better in an 
Euclidean approach: in a pioneering work~\cite{FHN}, published in 1975, 
Figari, H\o egh-Krohn and Nappi constructed the Wightman $n$-point functions of the $\mathscr{P}(\varphi)_2$ model
in a wedge of de Sitter space-time from the Schwinger functions  
on the Euclidean sphere. While the latter have explicit 
expressions in terms of Markov fields, the former are less accessible, as one is forced to rely on the intricate methods of 
analytically continuation in several variables.  

In this letter we will show that a direct, Lorentzian, non-perturbative approach based on group theoretic methods is feasible. 
For the \emph{free} theory, the representation theory of the isometry group $SO_0(1,2)$
provides the one-particle Hilbert space and the generators of the one-parameter subgroups implementing the boosts and the rotations. 
Second quantisation can be used to define \emph{canonical fields} (and the associated algebras of observables) on the 
corresponding \emph{Fock space}. The generators of the one-parameter subgroups 
on Fock space can be expressed in terms of the relevant conserved energy-momentum currents.  
Now, given the observables and the action of the symmetry group,  
one may try to find a replacement for the ``state of minimal energy''. In fact,  
the \emph{de Sitter vacuum state} for the free theory can be characterised by several criteria which all 
yield the same result:  the de Sitter  vacuum state for the free theory is induced by the Fock vacuum vector.

Our main result is that in $1+1$ dimensions,  the scalar bosonic
quantum field with polynomial interaction can \emph{as well} be formulated on Fock space, and understood  
by exploiting group theoretic methods: 
just as in classical field theory, explicit expressions for the conserved 
quantities (\emph{e.g.}, the generators of the interacting boosts) of the quantum theory 
\color{black} arise from the space-time symmetries, due to the Noe\-ther theorem. 
%And although Euclidean methods seem to be indispensable in order to establish certain aspects of the interacting 
%quantum theory on de Sitter space in full mathematical rigour,  much less is needed in order to come up with the 
%correct expressions for the interacting ${\mathscr P}(\varphi)_2$ model on de Sitter space. In fact, we find 
%it reassuring that the interacting quantum theory can be tailored along the lines of the classical theory, once one 
%has overcome some of the 
%conceptual hurdles which hampered progress on interacting quantum
%fields on curved space-time  
%in the past. 
After all, 
the~${\mathscr P}(\varphi)_2$ model on de Sitter space is, in our opinion, the simplest relativistic quantum field theory
that exists (in the strict mathematical sense), 
satisfying all the basic expectations one may have, such as finite speed of light, particle production, 
causality, stability of the vacuum and so on. It is also, by far the most explicit.  
As such, it is a pleasant surprise that this model can be presented on a few pages. 

With this motivation in mind, we dedicate Section \ref{sec-2} to a review of the classical situation. 
In Section \ref{Sect: canon-HS} we present a new formulation of the free theory (first derived in \cite{BJM-1}),
which makes it easy to add an interaction in the final section. 

\section{Classical Fields on De Sitter Space}
\label{sec-2}

In two-dimensions, de Sitter space $dS$
can be viewed as a one-sheeted \emph{hyperboloid}\index{hyperboloid}, embedded in 
three-dimensional Minkowski 
space~$\mathbb{R}^{1+2}$: 
\label{dSpage}
	\begin{equation}
		\label{eqdSMin}
		dS \doteq \left\{  
		x  \in \mathbb{R}^{1+2} 
		\mid 
		x_{0}^{2} - x_{1}^{2} - x_{2}^{2} = - r^2 \right\} \; . 
	\end{equation}
A convenient coordinate system, which covers the whole de Sitter space, is  given by
	\begin{equation}
	\label{psi-x0}
		\left(\begin{matrix} 
			x_0 	\\
			x_1	\\
			x_2
		\end{matrix}\right) 
		= 
		\left(\begin{matrix} 
			x_0 \\ 
			\sqrt{ r^2 + x_0^2 } \sin \psi \\ 
			\sqrt{ r^2 + x_0^2 } \cos \psi
		\end{matrix}\right)\; , \quad x_0 \in \mathbb{R} \; , \; \; \psi \in [0, 2 \pi) \; . 
	\end{equation} 
Expressed in the variables $x_0$ and $\psi$, the metric takes the form 
	\[
		g = \frac{r^2 }{r^2+x_0^2} \; {\rm d} x_0 \otimes {\rm d} x_0 
			- ( r^2+ x_0^2) \; {\rm d}  \psi \otimes {\rm d} \psi  \; . 
	\]
In this letter, we consider a classical {\em Lagrangian density}~${\mathcal L}(\mathbb{\Phi})$, which 
we view as a 2-form,  
	\begin{equation}
	\label{LagDen}
		{\mathcal L}(\mathbb{\Phi}) =  \frac{1}{2}  d \mathbb{\Phi} \wedge * d \mathbb{\Phi}  
		- \frac{\mu^2}{2} \mathbb{\Phi} * \mathbb{\Phi}  - \mathscr{P} ( \mathbb{\Phi})  * 1   \; . 
	\end{equation}
\color{black}
Here $\mathbb{\Phi}$ is a real valued scalar field, the star denotes the Hodge $*$ operator, 
the polynomial $\mathscr{P}$ is bounded from below and $\mu $ is a positive constant. 
The ensuing Euler-Lagrange equation is the  inhomogeneous Klein--Gordon equation: 
\label{squarepage} 
	\begin{equation}
		\label{3.25}
		(\square_{dS}+  \mu^2  ) \mathbb{\Phi}  = -
                \mathscr{P}' (\mathbb{\Phi} ) \; , 
		\quad \mathbb{\Phi} \in C^\infty (dS) \;  ,
	\end{equation}
where $\mu>0$ is the constant appearing in \eqref{LagDen}.  
The Noether Theorem associates with every vector field $X$ a conserved current $* \, \mathcal{T}_X$: 
\[
		d \left[  L_X \mathbb{\Phi} \wedge \tfrac{\partial {\mathcal L}}{\partial (d \mathbb{\Phi})} 
		- i_X {\mathcal L} \right] =:  d * \mathcal{T}_X = 0 \; . 
	\]
Using the coordinates 
	\[
		x (t,  \psi) = \underbrace{ \begin{pmatrix}
				 \cosh t  &  0 &\sinh t \cr
     					  0  &  1 & 0  \cr
 				  \sinh t &  0 & \cosh t \end{pmatrix} }_{ \equiv \Lambda_1(t)}
				\left( \begin{array}{c}
						0 \\
						r \sin\psi  \\
						r \cos\psi   
				\end{array} \right) \; . 
	\]
        the restriction 
        of the Killing vector field for the boosts 
	\[
		t \mapsto \Lambda_1(t)
	\]
on $dS$ to the circle $S^1= \{ x \in dS \mid x_0 =0 \}$ is given by $\partial_t $. 
If one integrates $* \mathcal{T}_{\partial_t}$ (respectively, $* \mathcal{T}_{\partial_\psi}$) 
over the space-like surface $S^1$,  then one finds the conserved quantities 
	\[
		\mathbb{L} = \int_{S^1} *  \mathcal{T}_{\partial_t}\quad \text{and} \quad
		\mathbb{K} = \int_{S^1} * \mathcal{T}_{\partial_\psi}   \; .
	\]
They generate the $\Lambda_1$-boosts 
and the rotations around the $x_0$-axis, respectively. 

Using the coordinates introduced in \eqref{psi-x0}, the conserved currents 
	\[
		\mathcal{T}_0 \equiv \mathcal{T}_{\partial_{x_0}} \; , \qquad \mathcal{T}_1 \equiv \mathcal{T}_{\partial_\psi}  \; , 
	\]
can be expressed in terms of \color{black} the \emph{classical stress-energy tensor} ${\mathbb{T}^\mu}_\nu$ defined by 
	\[
		{\mathcal{T}_\mu} = \mathbb{T}_{\mu\nu} dx^\nu \; ,
                \qquad x^0 \equiv t, \quad x^1 \equiv \psi \; . 
	\]
${\mathbb{T}^\mu}_\nu$ describes the flux of the $\mu$-th component of the conserved energy-momentum 
vector across a surface with constant $x^\nu$ coordinate. It is given by
\color{black}
	\begin{align*}
	{\mathbb{T}^{\mu}}_{\nu} & = \partial^\mu \mathbb{\Phi} \partial_\nu \mathbb{\Phi}  -  g^{\mu\kappa} 
	g_{\kappa \nu}{\mathcal L}(\mathbb{\Phi}) \\
        &= \partial^\mu \mathbb{\Phi} \partial_\nu \mathbb{\Phi} - \tfrac{1}{2} {\delta^{\mu}}_{\nu} 
	\bigl(  \partial^\kappa \mathbb{\Phi}\partial_\kappa \mathbb{\Phi} \bigr)
	+  {\delta^{\mu}}_{\nu}  \bigl(\tfrac{\mu^2}{2} \mathbb{\Phi}^2  + P (\mathbb{\Phi}) \bigr) .   
	\end{align*}

Rewriting $\partial_t\mathbb{\Phi}$ as 
	\begin{equation}
	\label{partial-t}
		\partial_t\mathbb{\Phi}= r\cos\psi \,  \partial_{x_0} \mathbb{\Phi} \equiv r\cos\psi \, \mathbb{\pi} \; , 
	\end{equation}
where $\partial_{x_0}$  is the future directed normal vector field 
restricted to the time-circle $x_0 =0$, we find
	\begin{align*}
		{\mathbb{T}^0}_{0} 
		&= \frac{1}{2} \left( \mathbb{\pi}^2 +   r^{-2} \bigl( \partial_\psi  \mathbb{\Phi} \bigr)^2 + 
					\mu^2 \mathbb{\Phi}^2 \right)  + P  (\mathbb{\Phi} ) \; ,  
	\end{align*}
with $\mathbb{\pi} =  \partial_{x_0}  \mathbb{\Phi}$, and 
	\begin{equation}
	\label{T01}
		{\mathbb{T}^{1}}_{0} 
		=   r^{-2} \partial_{x_0} \mathbb{\Phi}\, \partial_\psi \mathbb{\Phi} \; 
		= r^{-2}  \mathbb{\pi} \, \partial_\psi \mathbb{\Phi} \; . 
	\end{equation}
Integrating $\mathbb{T}_{00}$ over the time-zero circle~$S^1$ yields 
(after possibly adding a constant term 
to the interaction term $\mathscr{P}  (\mathbb{\Phi}$ ) to ensure $\mathscr{P}  (\mathbb{\Phi} )\ge 0$) a positive quantity, 
	\begin{equation}
	\label{T02}
		\int_{S^1} r\, {\rm d} \psi \; \mathbb{T}_{00}   (\psi) > 0 \; ,  
	\end{equation}
which may be interpreted as the \emph{total energy} for 
the classical $\mathscr{P}(\varphi)_2$ model on the \emph{Einstein universe} \cite{Fe1},\cite{Fe2}, \emph{i.e.}, 
the space-time of the form $S^1 \times \mathbb{R}$ with the metric induced from the ambient Minkowski space $\mathbb{R}^{1+2}$. 
However, the quantity on the left hand side 
of the inequality  in \eqref{T02} is \emph{not} a conserved quantity on de Sitter space and therefore should be disregarded. 

In contrast, using \eqref{partial-t} to replace the vector field $\partial_{x_0}$ by $\partial_{t}$ yields an explicit expression for the generator of the boost:
	\begin{align}
		\mathbb{L} =  
		\int_{S^1} r^2 \, \cos\psi \; {\rm d} \psi \; \mathbb{T}_{00} (\psi) \; . 
	\end{align}
The corresponding expression for the angular momentum is 
	\[ 
		\mathbb{K} = 
		\int_{S^1} \,  r  {\rm d} \psi \;   \mathbb{T}_{10} 
		=  \int_{S^1} \,  r \, {\rm d} \psi \;   \mathbb{\pi} \, (\partial_\psi \mathbb{\Phi})  \; .
	\]
We will soon encounter very similar formulas for the quantum fields.

\section{Canonical Free Quantum Fields}
\label{Sect: canon-HS}

We will now provide a description of the quantum theory on two-dimensional de Sitter space, 
which follows closely the classical theory exposed in the previous section.  
We start by discussing the one-particle theory, following Wigner's prescription to identify 
the one-particle space with the Hilbert space carrying the unitary irreducible representation of the 
symmetry group. The single-particle wave-functions 
on the Cauchy surface $S^1 = \{ x \in dS \mid x_0 =0 \}$ can be viewed as elements of   
	\[
		\mathcal{H} \doteq L^2 (S^1, r {\rm d} \psi) \;  .  
	\]
The Hilbert space $\mathcal{H}$ carries \cite{BJM-1} a unitary irreducible representation of $SO_0(1,2)$ 
for $\mu>0$ and spin zero, generated by the rotations 
	\[
	\bigr(\widehat{u} (R_0(\alpha)) h \bigl) (\psi) 
	= h (\psi - \alpha) \; , \quad \alpha \in [0, 2\pi) \; , \; h \in \mathcal{H} \; , 
	\]
and the boosts   
	\begin{equation} 
		\label{eqUmhat}
			\widehat{u} (\Lambda_1(t)) 
			= {\rm e}^{i t \sqrt{ \omega } \, r \mathbb{cos} \sqrt{\omega }  } \; , \qquad t \in \mathbb{R} \; . 
	\end{equation}
The symbol $\mathbb{cos}$ denotes the multiplication operator 
	\[
		(\mathbb{cos} \; g) (\psi) = \cos \psi \cdot g (\psi) \; , \qquad g \in \mathcal{H} \; . 
	\]
The Fourier coefficients $\widetilde {\omega}(k) = \widetilde {\omega}(- k)$, $k \in \mathbb{Z}$, of the strictly positive 
self-adjoint operator $\omega $ appearing in \eqref{eqUmhat} are strictly positive. They can be expressed in terms 
of~$\Gamma$~functions:  
	\begin{align}
	\label{eq:omega-1}
		\widetilde {\omega}(k)  & =  \frac{k+s^+}{r} \, 
					 \frac{\Gamma \left( \frac{k+s^+}{2} \right)}{ \Gamma \left( \frac{k-s^+}{2} \right)}
			\frac{ \Gamma \left( \frac{k+1-s^+}{2} \right) }{ \Gamma \left( \frac{k+1+s^+}{2} \right)} \; .
	\end{align}
The constant $\mu   >0$ enters only through the parameter~$s^+$.  The latter is given by 
	\begin{equation} 
		\label{dd1} 
			s^\pm= -\frac{1}{2}  \mp i \nu \; , \; \;  \nu =  
			\begin{cases}
				i \sqrt{\frac{1}{4} -   \mu^2  r^2} \, , &  0<   \mu  < \tfrac{1}{2r}   \, ,\\
				\sqrt{  \mu^2  r^2 - \tfrac{1}{4}} \, , &   \mu  \ge \tfrac{1}{2r}  \, .
			\end{cases} 
	\end{equation} 
For $|k|$ large, 
	\[ 
		\omega (k) \sim \sqrt{\tfrac{k^2}{{r}^2}  +  \mu^2 } \; , \qquad 1 \ll k \; . 
	\]
We note that if the time-zero circle $S^1$ were the Cauchy surface of the Einstein 
cylinder $S^1 \times \mathbb{R}$ (instead of $dS$), 
the operator $\omega r $ would be the generator of the free time evolution. The generator of the boost, 
$\sqrt{ \omega } \, r \mathbb{cos} \sqrt{\omega }$,  differs from $\omega r$ by 
the inserted cosine function which takes care of the length of the Killing vectors for the boost 
on~$dS$ at~$x_0=0$. 

Given the one-particle description, we may proceed by \emph{second quantisation}. As the Fock space, we choose 
	\[
		\Gamma ( \mathcal{H} )  \doteq\oplus_{n= 0}^{\infty}  \Gamma^{(n)} ( \mathcal{H} ) \;  , 
	\]
with $ \Gamma^{(0)} ( \mathcal{H} ) \doteq \mathbb{C}$ and $ \Gamma^{(n)} ( \mathcal{H} ) $, $n \in \mathbb{N}$, 
the $n$-fold totally symmetric tensor product $\otimes_s$ of~$\mathcal{H} \doteq L^2 (S^1, {\rm d} \psi)$ with itself. 

The \emph{canonical time-zero fields} and the \emph{canonical momenta} 
are expressed in terms of creation and annihilation operators:
	\begin{align*} 
		\varphi(\psi) & =  \frac{1}{\sqrt{2}}
						\Big(\big(\omega^{-\frac{1}{2}} a\big)(\psi)^* + \big(\omega^{-\frac{1}{2}} a\big)(\psi)\Big)  \;,   \\
		\pi(\psi) & =  \frac{i}{\sqrt{2}}   \Big(\big(  \omega^{\frac{1}{2}}  a\big)(\psi)^* - \big(  \omega^{\frac{1}{2}}  
		a\big)(\psi)\Big) \; , 
	\end{align*}
where the creation and annihilation operators satisfy
	\begin{equation*}
		\big[a(\psi'), \; a^*(\psi)\big]   =  \tfrac{1}{r} \delta(\psi-\psi')   
        \end{equation*}
as well as $\big[a(\psi')^*, \; a(\psi)^*\big]=\big[a(\psi'), \; a(\psi)\big]=0$. 
The canonical fields and momenta satisfy \emph{canonical commutation relations}. In particular, 
	\[
		\big[\pi(\psi'), \; \varphi(\psi)\big]=-   \frac{i}{r}  \delta(\psi-\psi') \; .
	\]
        
A unitary representation $\Lambda \mapsto U_\circ (\Lambda)$ of $SO_0(1,2)$ on the 
Fock space is provided by second quantisation as well: the generators of the free boosts (leaving~$W_1$ invariant) and 
the free rotations on $\Gamma (\mathcal{H})$ are 
	\begin{align*}
		L_\circ  \doteq {\rm d}\Gamma 
		\bigl( \sqrt{ \omega } \, r \mathbb{cos} \sqrt{\omega }  \bigr) \; , 
		\quad 
		K_\circ   \doteq {\rm d} \Gamma 
		\bigl( - i \partial_\psi  \bigr) \; , 
	\end{align*}
respectively. Expressing $K_\circ$ in terms of 
canonical fields and canonical momenta, a straight forward (but rather lengthy) computation 
shows that 
	\[
		K_\circ = \int_{S^1} \,   
		r \,  {\rm d} \psi \;  T_{10}  \; ,
	\qquad
	\text{with}
	\qquad
				T_{10} = {:} \, \pi \,
                 \frac{\partial \varphi}{\partial \psi} (\psi) \,  {:} \,   . 
	\]
This expression is of the same form as the classical expression given by \eqref{T01}.

\goodbreak
We can now define the \emph{Wightman $n$-point functions} for $n$ arbitrary points in~$dS$:
	\[
		\mathcal{W}^{(n)} (x_1, \ldots , x_n) \doteq \langle \Omega_\circ , \Phi (x_1) \cdots \Phi (x_n) \Omega_\circ \rangle  \; , 
	\]
with $x_i = \Lambda_i \left( \begin{smallmatrix} 0\\0\\  r   \end{smallmatrix} \right)$, $i= 1, \ldots ,n$, and
	\[
		\Phi (x_i) \doteq U_\circ (\Lambda_i) \varphi (0) U_\circ^{-1} (\Lambda_i) \; , \quad 
		\Lambda_i \in SO_0(1,2)  \; . 
	\]
Note that this definition is consistent, as
in the sense of quadratic forms we have \cite{BJM-1} 
	\[
		U_\circ (\Lambda_2 (t)) \varphi (0) U_\circ^{-1} (\Lambda_2(t)) = \varphi (0)   \quad 
		\forall t \in \mathbb{R} \; . 
	\]

It remains to justify that the Fock zero-particle vector $\Omega_\circ$ induces the 
\emph{physically relevant de Sitter vacuum state}. This is not completely obvious: as mentioned before, 
there is no global time evolution on de Sitter space (in terms of a one-parameter group of isomorphisms) and hence
no natural notion of energy. Consequently, one can not require that the \emph{de Sitter vacuum state} 
is a state of minimal energy. One may still require that a  \emph{de Sitter vacuum state} is invariant under the action 
of the Lorentz group. But in itself, this requirement does \emph{not} guarantee the necessary stability properties. 

Stability of the vacuum state is a basic condition that can not simply be abandoned. It includes
stability of matter against spontaneous collapse, and it also stipulates 
that the energy-momentum currents do not fluctuate in an uncontrollable manner. It thereby excludes the 
so-called $\alpha$-vacua \cite{BFH}. The \emph{geodesic KMS condition} proposed by Borchers and Buchholz 
ensures such stability properties: it requires that the restriction of the vacuum state to the wedge\footnote{Due to Lorentz invariance, 
the geodesic KMS condition applies to any wedge $W= \Lambda W_1$, $\Lambda \in SO_0(1,2)$, with 
	the relevant boosts $t \mapsto \Lambda_{\scriptscriptstyle W}(t)$ now given by 
	$\Lambda_{\scriptscriptstyle W}(t) \doteq \Lambda \Lambda_1(t) \Lambda^{-1} $,  $t \in \mathbb{R}$.} 
	\[
		W _1\doteq \bigl\{  x \in dS \mid x_2 > |x_0 | \bigr\} \; , 
	\]
is a thermal state with respect to the ``dynamics'' provided by the one-parameter group $t \mapsto U_\circ (\Lambda_1(t))$ 
of boosts which leaves the wedge $W_1$ invariant.

The zero-particle vector $\Omega_\circ$ induces the \emph{unique}\footnote{Uniqueness of the vacuum state for free scalar field 
on the de Sitter space has been established by several authors 
using different but equivalent characterisations of the vacuum. If one uses the geodesic KMS condition to characterise the vacuum state, 
this very condition fixes the free $2$-point function. Uniqueness of the interacting vacuum state follows from operator algebraic arguments, 
see \cite{BJM-1} for details.} Lorentz invariant state 
(not only as Fock vector but as a state on the algebra) that satisfies the geodesic KMS condition. 
For free fields, the vector $\Omega_\circ$ induces also the unique Lorentz invariant state, which satisfies the 
so-called \emph{Hadamard condition}. The latter requires that the short distance behaviour of the two-point function 
is equal to that of the two-point function on Minkowski space.  
It was reformulated (and renamed as \emph{microlocal spectrum condition}) by Radzikowski \cite{Raz2} as 
a requirement for the \emph{wave front set} of the two-point function. 
The latter is nowadays widely used for non-interacting theories on more general curved space-times. However, 
the advantage of the geodesic KMS condition (which explicitly refers to de Sitter space)
is that it remain meaningful for interacting theories. 

\section{Interacting Quantum Fields}

In de Sitter space the spatial volume is compact. 
Thus Haag's theorem does not apply, and one may hope that it is possible to 
describe the interacting theory on Fock space. This is in fact possible, as
the ultraviolet problems are tame in~$1+1$ space-time dimensions.
As a consequence, the interacting relativistic quantum field theory constructed in this 
section is the most explicit and most accessible we have ever encountered. 

If the action of the subgroup preserving a geodesic Cauchy surface is identical 
to the one in the free theory, 
the interacting quantum fields differ from the canonical free fields only by the 
action of the interacting boosts. Given a new representation $U$ of $SO_0(1,2)$ 
(called the interacting representation), representing the interacting dynamics, we 
intend to proceed as in the non-interacting case by setting 
	\[
		\Phi_{int} (x) \doteq U(\Lambda) \varphi (0) U^{-1} (\Lambda) \; , \quad 
		x = \Lambda \left( \begin{smallmatrix} 0\\0\\ r  
		\end{smallmatrix} \right)  \; . 
	\]
Note that the boost $t \mapsto \Lambda_2(t)$ leave the point 
$\left( \begin{smallmatrix} 0\\0\\ r  \end{smallmatrix} \right)$ invariant.  Thus, 
in order to guarantee that there is no ambiguity in this definition, we will have to ensure that 
	\begin{equation}
	\label{UL2}
		U (\Lambda_2 (t)) \varphi (0) U^{-1} (\Lambda_2(t)) = \varphi (0)   \quad 
		\forall t \in \mathbb{R} \; .  
	\end{equation}
As soon as we have specified $U (\Lambda_2 (t))$, we will return to this point. 
Since the rotations and the boosts $t \mapsto \Lambda_1 (t)$  
generate $SO_0(1,2)$ it is sufficient to provide explicit expressions for the generators 
replacing $L_\circ $ and  $K_\circ$. In fact, there is no need to change the generator
of the rotations.  Just as in the free case, it is given by $K_\circ$. 
Looking at the classical analog, one may expect that the generator 
	\[
			L =  \int_{S^1} \,  r^2 \cos  \psi    \, {\rm d} \psi  \; T_{00}  (\psi) \; , 
	\]
of the interacting boost is given by integrating the energy density
${{\rm T}^0}_{0}  (\psi)$ against the cosine function.  As in the free case, 
the {\em energy density\/} ${{\rm T}^0}_{0} (\psi)$ is the restriction of the energy density 
% in the time-zero plane (in the ambient Minkowski space) 
to the Cauchy surface $S^1$, \emph{i.e.},   
	\begin{align}
	\label{energymomentumdensity2}
		 {T^0}_{0}   (\psi) =  \tfrac{1}{2}  \bigl(: \pi(\psi)^2:
						+ \tfrac{1}{r^2}
                                             :\tfrac{
                                                  \partial \varphi
                                                }{\partial \psi} (\psi)^2:  
						 + \mu^2 :\varphi(\psi)^2:  \bigr)
						  + {:} \mathscr{P} (\varphi(\psi)) {:} \; .  
	\end{align}
%In case the interaction $\mathscr{P}$  vanishes, $L$ is equal to $L_\circ$. 
Moreover, the structure of the sum in \eqref{energymomentumdensity2}
ensures that one can use the Trotter product formula to show that~\eqref{UL2}
holds. Euclidean methods were used in \cite{BJM-1} to  show that $K_\circ$ and $L$ are 
self-adjoint operators on Fock space, and that they generate
one-parameter subgroups which give rise to a new (in 
case ${:} \mathscr{P} (\varphi(\psi)) {:} \ne 0$) representation of the Lorentz group acting on Fock space, just as 
we have tactically assumed at the beginning of this section. 

\goodbreak
It is important to note that 
the interacting quantum field ${\Phi}_{\rm int} ( x ) $, $x \in dS$,   satisfies the {\em covariant equation of motion}  
on Fock space:  
	\[
		\Bigl( \square_{dS}+\mu^2 \Bigr) {\Phi}_{\rm int} ( x ) =   - {:} {\mathscr P}' ( {\Phi}_{\rm int} ( x )){:} \; .
	\]
This result may be verified by explicit computation, using the fact that the $\square_{dS}$ can be expressed in terms of the generators 
of the boosts and the rotations; see~\cite{BJM-1} for details.
Similar equations of motion for interacting quantum fields on Minkowski space were derived by Glimm and  Jaffe \cite{GJ2},  
Schrader \cite{Sch} and, in $2+1$ space-time dimensions, by Feldman and~Raczka \cite{FR}. 
However, in the case studied in the literature the equation of motion can \emph{not} be realised in Fock space 
(due to Haag's theorem), and the meaning of the right hand side is less evident. 

The de Sitter vacuum state for the ${\mathscr P}(\varphi)_2$ model can again be 
characterised by the geo\-desic KMS condition.  As it turns out, it  
has some surprising properties. Due to the thermalisation effects introduced by the curvature of space-time, 
it is \emph{unique} even for large coupling  constants, despite the fact that different phases occur in the limit of curvature to zero (i.e., 
the Minkowskian limit). As the ultraviolet problems are tame in $1+1$ space-time dimensions and
the spatial volume is compact, the \emph{interacting de Sitter vacuum state} is induced  
by a vector in Fock space:
	\begin{equation}
	\label{int-vac}
		\Omega = \frac{{\rm e}^{-\pi H}\Omega_\circ }{ 
		\|{\rm e}^{-\pi H}\Omega_\circ  \| } \; , 
	\end{equation}
where
	\[
	 	H := L_\circ + \int_{0}^{\pi}  r^2  \cos  \psi  \, {\rm d} \psi \;   {:} {\mathscr P}(\varphi(\psi) {:}   \; . 
	\] 
The formula \eqref{int-vac} is well known from the perturbation theory of thermal equilibrium states. By construction, 
we have $L \Omega =0$. What is less evident is that the vector $\Omega$   
is also invariant under rotations. This can be seen by using a Feynman-Kac formula 
to define a rotation invariant Euclidean vacuum state on the Euclidean sphere 
starting from \eqref{int-vac}, see \cite{BJM-1}.

\section{Summary} 

We have constructed the Fock space associated to the geodesic Cauchy surface; thus in contrast to many other works
in the literature, 
our description covers the whole de Sitter space and not just the region accessible to a single observer. We have
shown that the interacting field is well-defined, as an operator valued distribution acting on the Fock space 
of the free massive scalar field, as it differs only by the adjoint action of a unitary operator from the free time-zero field. 
We have stated the equation of motion, which the interacting field satisfies, and we have
provided explicit expressions for the generators of the boosts and the rotations, generating a representation of the space-time
symmetry group, in terms of (normal ordered) free time-zero fields (and their 
derivatives), which are similar to the corresponding classical expressions provided by the Noether theorem. Finally, 
we have also provided explicit formulas for the de Sitter vacuum states, which in both the free and the interacting case 
are induced by vectors in Fock space, and which in both cases 
enjoy stability properties that make them unique. 

\paragraph*{Acknowledgments.}
Jens Mund was partially supported by the São Paulo Research Foundation (FAPESP), grant number 2014/24522-9, 
and both authors were supported by the Conselho Nacional de Desenvolvimento Cientifico e Tecnológico (CNPq).
We would like to thank Karl-Henning Rehren and Urs Wiedemann for critical comments on an earlier version of this letter.

\end{document}